\documentclass{article}
\usepackage{amsmath}
\usepackage{amssymb}
\usepackage{color}
\usepackage{authblk}

\usepackage{vmargin}                   
\setmarginsrb{1.5cm}{1.5cm}{1.5cm}{1.5cm}{0cm}{1cm}{0cm}{1cm}   

\usepackage{graphicx}
\usepackage{bm}

\newcommand\Tt{\mathbf{T}}

\newcommand\rr{{\bf r}}
\newcommand\ee{{\bf e}}
\author[1]{Vincent Tejedor}
\affil[1]{Cour des comptes, 13, rue Cambon, 75 001 Paris, France}

\begin{document}
\title{First passage properties in a crowded environment}

\maketitle

\section*{Introduction}

\begin{figure}[htb]
\centering \includegraphics[width =0.5\linewidth,clip]{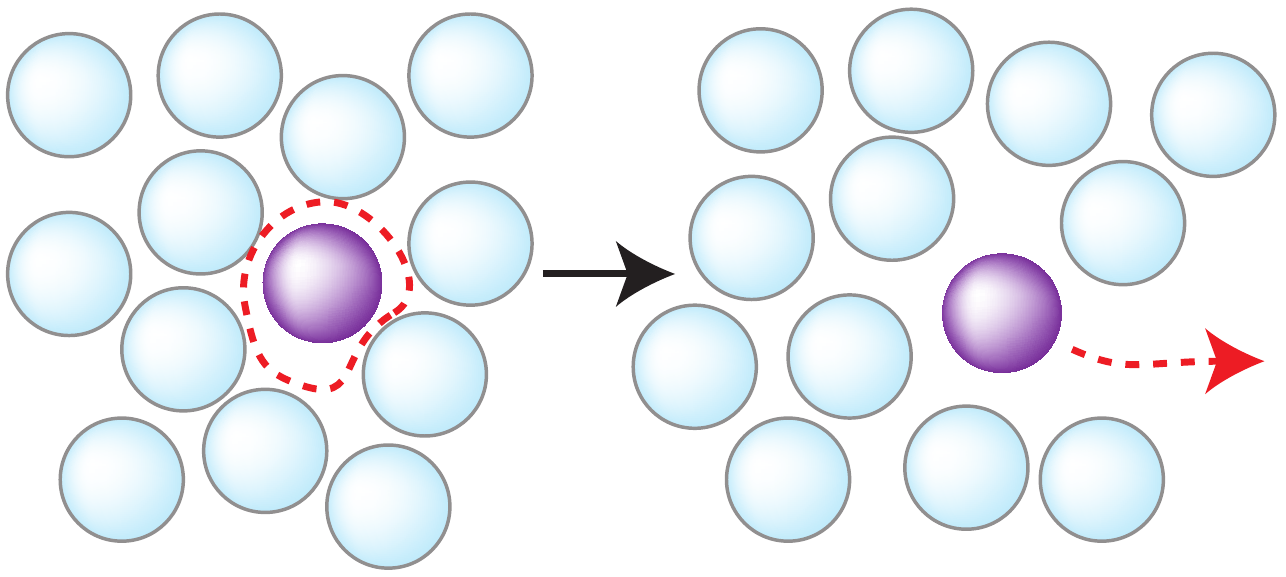} 
\caption{(color online) Crowding and dynamical cages: surrounding particles can form, for a given time, a ``cage'' around the diffusing molecule.}
\label{fig:r2Cube}
\end{figure}

Crowding effects are often pointed as a possible cause of the anomalous diffusion observe within a cell \cite{Saxton:1996,Saxton:2007}. The crowding obstacles form dynamical cages around a tracer \cite{Condamin:2008}. This cage disappears after some time, releasing the tracer, as shown in figure \ref{fig:r2Cube}. This model leads to a continuous time random walk (CTRW) behavior, with a natural cut-off of the waiting time distribution, somehow related to the obstacle density. In experimental trajectories, such as lipid granules diffusing in a fission yeast cell, it has been observed that at short time, the diffusive behavior shared a lot of common features with CTRW \cite{Tejedor:2010kx}. 

We will here develop a model on an Euclidian lattice, to investigate the crowding influence on the mean first passage time (MFPT), and more generally on first-passage properties. To obtain the MFPT, we will first consider the problem with only one vacancy (over-crowded case) and solve it exactly, before extending, with an approximation, this result to an arbitrary number of vacancies.

\section{Model}

We will consider a crowding problem on a discrete network. This allow somehow a simplification of the problem, since the crowding effect can be simply defined as follows: two particles can not occupy the same node.

On the considered network, we will mix one or several obstacles, each occupying one node, and a tracer particle, also occupying one node. At each time step, all obstacles will successively choose a neighboring site, and move toward it if the site is free. Since the obstacles are all the same, the movement order does not really matter. After the obstacle diffusion, the tracer chooses a neighboring site, and move toward it if there is no obstacle already occupying this node. Else, the tracer stays in the same position.

We can see the problem symmetrically by considering the vacancies: instead of considering $n_{\rm o}$ obstacles, we can assume that the network contains $n = V-(n_{\rm o}+1)$ vacancies, where $V$ is the network volume, {\it i.e.} the number of site, and the $+1$ term comes from the tracer. Each vacancy is an independent random walker, and two vacancies can not occupy the same site. In this vision, the tracer moves as soon as a vacancy hit him: the tracer position and the vacancy position are simply exchanged. As previously, the vacancies move first, and if a vacancy hit the tracer, the tracer moves. At last, we impose that the tracer can only move once at each time step.

\bigskip

To allow an analytical treatment, the network considered will be a periodic Euclidian lattice in $d$ dimensions, the lattice being $X_i$ large in the $i$ direction. 

\break

\subsection{Problem for a single vacancy}

We will start with a single vacancy, $n=1$. This problem can be solved exactly for certain geometries. Our goal will be to obtain the mean first-passage time $\langle \Tt_t \rangle (\rr_T,\rr_S)$ for the tracer between an initial position $\rr_S$, and a target site $\rr_T$.

At each step, the tracer has to wait that the vacancy comes back, and thus performs a continuous time random walk. The waiting time is the time needed for the vacancy to hit again the tracer. Since the tracer motion is related to the path chosen by the vacancy to hit the tracer, the latter performs an anti-persistent random walk: it is more likely that the vacancy hits the tracer from the opposite direction of the last step than in the same direction.

\bigskip

To obtain the MFPT, we will use the MFPT of a persistent random walk: the tracer performs indeed an anti-persistent CTRW. We will first compute the persistence probabilities of the random walk for the single vacancy problem (a), and then the mean waiting time between two steps (b). Assembling those two results, we will deduce the mean first-passage time of a tracer in a discrete network with a single vacancy (c).

\subsubsection{Conditional step probabilities}

As shown previously, to obtain the MFPT for a persistent random walker, one has first to compute the conditional step probabilities, namely the probability to perform a step in a given direction, knowing what was the last direction chosen.

We will consider an Euclidian lattice of dimension $d$, where $(\ee_1, \ldots, \ee_d)$ is an orthonormal base. If we note $\rr(t)$ the position of the tracer at time $t$, $\rr_V (t)$ the position of the vacancy, making a step in the direction $\ee_i$ at time $t$ means that $\rr(t+1) = \rr(t) + \ee_i$ and $\rr_V(t+1) = \rr(t)$. The vacancy then starts from $\rr(t)$, and the next step occurs at $t'>t$ when $\rr_V(t') = \rr(t+1)$. The tracer will go in $\rr_V(t'-1)$: to compute the probability to go in the direction $\ee_j$, we have to compute the probability that $\rr_V(t'-1) = \rr(t+1) + \ee_j$ if $\rr_V(t') = \rr(t+1)$ for the first time since $t$, knowing that $\rr_V(t+1) = \rr(t+1) - \ee_i$. We have here a random walk with a short memory: the probability to make a step in direction $\ee_j$ depends on the position $\rr(t)$, but also on the direction $\ee_i$ of the last step. We can define the conditional probability to perform a step in direction $\ee_j$ after a step in direction $\ee_i$ starting from $\rr(t)$:
\begin{equation}
{\rm Prob}(\ee_j|\ee_i, \rr(t) ) = {\rm Prob}\Big ( \rr_V(t'-1) = \rr(t)+\ee_j | \rr_V(t) = \rr(t)-\ee_i \Big  ),
\end{equation}
where $t'$ is defined as:
\begin{equation}
t' = \min \left ( t_1 \in \mathbb{N} \big / t_1 > t \ \&  \ \rr_V(t_1) = \rr (t) \right )
\end{equation}
$t'$ is here the first-passage time of the vacancy to the initial tracer position.

To compute this conditional probability, we need to know the last position of the vacancy before hitting the tracer, knowing that the vacancy start in $\rr(t)-\ee_i$. This problem can be formalized using an electrical analogy \cite{Condamin:2007eu}: we consider that we inject a flux $J$ on the network in $\rr(t)-\ee_i$, and we remove a flux $J$ in $\rr(t)$. Our problem is to determine the flux $J_{\pm j}$ on each edge $\rr(t) \pm \ee_j \to \rr(t)$. Using the pseudo-Green function, we can write that the stationary density (for the flux problem) fulfills:
\begin{equation}
\rho(\rr_i) = J \Big ( H \left ( \rr(t) | \rr(t) \right ) - H \left ( \rr(t) | \rr(t)-\ee_i \right )  +  H \left ( \rr_i | \rr(t)-\ee_i \right ) -  H \left ( \rr_i | \rr(t) \right ) \Big ),
\end{equation}
where $H(\rr_i | \rr_j)$ is the pseudo-Green function. As previously, this function is defined using the propagator $P \left ( \rr_i | \rr_j,t \right )$, probability to be in $\rr_i$ at time $t$, starting from $\rr_j$ at time $t=0$, and $P_{\rm \small stat}(\rr_i)$, the stationary probability (for the original network, without flux) to be in $\rr_i$:
\begin{equation}
H \left ( \rr_i | \rr_j \right ) = \sum_{t=0}^{\infty} \Big ( P \left ( \rr_i | \rr_j,t \right ) - P_{\rm stat}(\rr_i) \Big )
\end{equation}
We can thus compute the stationary probability in $\rr(t) \pm \ee_j$, and deduce the flux $J_{\pm j}$:
\begin{eqnarray}
\nonumber \rho(\rr(t)\pm \ee_j) & = & J \Big ( H \left ( \rr(t) | \rr(t) \right ) - H \left ( \rr(t) | \rr(t)-\ee_i \right ) \\
& & \hspace{2cm} +  H \left ( \rr(t)\pm \ee_j | \rr(t)-\ee_i \right ) -  H \left ( \rr(t)\pm \ee_j | \rr(t) \right ) \Big ),\\
J_{\pm j} & = & \omega_{\rr(t)\pm \ee_j \to \rr(t)} \rho(\rr(t)\pm \ee_j),
\end{eqnarray}
where $\omega_{\rr(t)\pm \ee_j \to \rr(t)}$ is the transition probability from $\rr(t)\pm \ee_j$ to $\rr(t)$. 

\break

We finally obtain the probability:
\begin{eqnarray}
{\rm Prob}\left (\ee_j|\ee_i, \rr(t) \right ) & = &\displaystyle \frac{J_{j}}{J}\\
& = &  \omega_{\rr(t) + \ee_j \to \rr(t)} \Big ( H \left ( \rr(t) | \rr(t) \right ) - H \left ( \rr(t) | \rr(t)-\ee_i \right ) \nonumber \\
& & \hspace{1cm} +  H \left ( \rr(t) + \ee_j | \rr(t)-\ee_i \right ) -  H \left ( \rr(t) + \ee_j | \rr(t) \right ) \Big )
\end{eqnarray}

Some comments are in order: 
\begin{itemize}
\item[(i)] Those probabilities are entirely determined by the vacancy motion, which is so far a classical Brownian motion. In particular, the pseudo-Green functions can be known exactly for some geometries. 
\item[(ii)] The tracer motion is persistent for the general case. One can easily understand that it is easier for the vacancy to come back using the same path than to circle around the tracer to arrive in the opposite direction: the tracer performs an anti-persistent motion. 
\item[(iii)] The transition probabilities for the tracer depend on the position $\rr(t)$ in the general case.
\end{itemize}

We can slightly simplify the problem if we consider a tracer moving on a periodic Euclidian lattice in $d$ dimensions. The network is parallepipedic, being $X_i$ long on direction $\ee_i$. For such network, the transition probability does not depends anymore on the tracer position, and the pseudo-Green function is exactly known:
\begin{equation}
H(\rr | \rr' ) = \frac{1}{\displaystyle \prod_{i=1}^d X_i} \sum_{{\bf q} \neq {\bf 0}} \frac{\displaystyle e^{2 \imath \pi {\bf q}.(\rr - \rr ' )}}{\displaystyle 1-\frac{1}{d} \sum_{i=1}^d \cos \left ( 2 \pi {\bf q}.\ee_i \right )},
\end{equation}
where ${\bf q} = (1/X_1,\ldots,1/X_d)$. We can note that the pseudo-Green function only depends on the difference $\rr-\rr'$: $H(\rr | \rr') = H(\rr-\rr')= H(\rr'-\rr)$. The tracer transition probability becomes:
\begin{eqnarray}
{\rm Prob}\left (\ee_j|\ee_i \right ) & = & \frac{1}{2d} \Big ( H \left ( {\bf 0} \right ) - H \left ( \ee_i \right ) +  H \left ( \ee_j + \ee_i \right ) -  H \left ( \ee_j\right ) \Big ) \nonumber \\
& = & \frac{1}{\displaystyle 2d \prod_{i=1}^d X_i} \sum_{{\bf q} \neq {\bf 0}} \frac{\displaystyle 1- e^{ 2 \imath \pi {\bf q}.\ee_i}+ e^{2 \imath \pi {\bf q}.(\ee_j+\ee_i)} - e^{2 \imath \pi {\bf q}.\ee_j}}{\displaystyle 1-\frac{1}{d} \sum_{i=1}^d \cos \left ( 2 \pi {\bf q}.\ee_i \right )} \label{eq:Probcond}
\end{eqnarray}

One can note that if $X = X_1=X_2=\ldots=X_d$ (hypercubic domain) we have only three probability, as soon as $d \geq 2$: Prob$(\ee_i,\ee_i)$, Prob$(\ee_i,-\ee_i)$ and Prob$(\ee_i,\ee_{j\neq i})$. We will note them respectively Prob$^{\rightarrow \rightarrow}$, Prob$^{\rightarrow \leftarrow}$ and Prob$^{\rightarrow \uparrow}$.

One could argue that the first return time is not the same for each couple $(\ee_i,\ee_j)$. The correct mean first-passage time $\langle \Tt_t \rangle$ for the tracer should be:
\begin{equation}
\langle \Tt_t (\rr_S,\rr_T) \rangle = \sum_{i=1}^{d}\sum_{j=1}^d \langle N  (\rr_S,\rr_T) \rangle ( \pm \ee_i, \pm \ee_j ) \langle \psi \rangle ( \pm \ee_i, \pm \ee_j ),
\end{equation}
where $\langle N \rangle (\ee_i, \ee_j )$ is the average number of reorientation from $\ee_j$ to $\ee_i$ during the tracer walk, and $\langle \psi \rangle ( \ee_i, \ee_j )$ the mean waiting time between a step along $\ee_j$ and a step along $\ee_i$. The two random variables $N$ and $\psi$ are uncorrelated, we can thus multiply the means to obtain the mean of the product. $\langle \psi \rangle ( \ee_i, \ee_j )$ is here a conditional mean first-passage time for the vacancy: starting from $\rr-\ee_j$, this is the mean first-passage time to $\rr$ when the last step occurs along $\rr-\ee_i \to \rr$. We will compute successively $\langle N \rangle (\ee_i, \ee_j )$ and $\langle \psi \rangle ( \ee_i, \ee_j )$, after some simplifications.

\bigskip

Assuming once again that the problem occurs in a hypercube of size $X$ in $d$ dimensions, we can slightly simplify the problem:
\begin{equation}
\langle \Tt_t (\rr_S,\rr_T) \rangle = \langle N^{\rightarrow \rightarrow} (\rr_S,\rr_T) \rangle  \langle \psi^{\rightarrow \rightarrow} \rangle + \langle N^{\rightarrow \leftarrow} (\rr_S,\rr_T) \rangle  \langle \psi^{\rightarrow \leftarrow} \rangle + \langle N^{\rightarrow \uparrow} (\rr_S,\rr_T) \rangle  \langle \psi^{\rightarrow \uparrow} \rangle.
\end{equation}

The problem can be further simplified. We note $N^{\rightarrow \uparrow}(\rr_T|\rr_S,\ee_i)$ the number of reorientation $\rightarrow \uparrow$ starting from $\rr$, the previous step being $\ee_i$ and the target $\rr_T$. 

\break

This quantity satisfies, for $\rr \neq \rr_T$:
\begin{eqnarray}
\langle N^{\rightarrow \uparrow}\rangle (\rr_T|\rr,\ee_i) & = & {\rm Prob}^{\rightarrow \rightarrow} \langle N^{\rightarrow \uparrow}\rangle(\rr_T|\rr-\ee_i,\ee_i) + {\rm Prob}^{\rightarrow \leftarrow} \langle N^{\rightarrow \uparrow} \rangle(\rr_T|\rr-\ee_i,-\ee_i) \nonumber \\
& & \hspace{2cm} + {\rm Prob}^{\rightarrow \uparrow} \sum_{\pm\ee_j, j \neq i}  \langle N^{\rightarrow \uparrow} \rangle(\rr_T|\rr-\ee_i,\ee_j) + {\rm Prob}^{\rightarrow \uparrow}.
\end{eqnarray}
We can compare it with the equation satisfied by the mean first-passage time for a ``classical'' persistent random walker, namely a persistent random walker waiting one time step between two jumps, starting from $\rr$ with orientation $\ee_i$, when $\rr_T$ is the target, $\langle \Tt \rangle(\rr_T|\rr,\ee_i)$, once again for $\rr \neq \rr_T$:
\begin{eqnarray}
\langle \Tt \rangle (\rr_T|\rr,\ee_i) & = &  {\rm Prob}^{\rightarrow \rightarrow} \langle \Tt \rangle (\rr_T|\rr-\ee_i,\ee_i) + {\rm Prob}^{\rightarrow \leftarrow} \langle \Tt \rangle (\rr_T|\rr-\ee_i,-\ee_i) \nonumber \\
& & \hspace{2cm} + {\rm Prob}^{\rightarrow \uparrow} \sum_{\pm \ee_j,j \neq i}  \langle \Tt \rangle (\rr_T|\rr-\ee_i,\ee_j) + 1. \label{eq:resp2}
\end{eqnarray}
Both quantities satisfy the same equation, except for the source term. We can then deduce that:
\begin{eqnarray}
\langle N^{\rightarrow \rightarrow} \rangle (\rr_T|\rr_S,\ee_S) & = & {\rm Prob}^{\rightarrow \rightarrow} \langle \Tt \rangle (\rr_T|\rr_S,\ee_S)\\
\langle N^{\rightarrow \leftarrow} \rangle (\rr_T|\rr_S,\ee_S) & = & {\rm Prob}^{\rightarrow \leftarrow} \langle \Tt \rangle (\rr_T|\rr_S,\ee_S)\\
\langle N^{\rightarrow \uparrow} \rangle(\rr_T|\rr_S,\ee_S) & = & {\rm Prob}^{\rightarrow \uparrow} \langle \Tt \rangle (\rr_T|\rr_S,\ee_S)
\end{eqnarray}
Those equations are still valid after an average over $\ee_S$. $\langle \Tt \rangle (\rr_T|\rr_S,\ee_S)$ is the mean first-passage time for a persistent random walker waiting a time unit between two steps, and is analytically known, using equation (\ref{eq:resp2}).

The mean first-passage time for our problem becomes:
\begin{equation}
\langle \Tt_t (\rr_S,\rr_T) \rangle = \langle \Tt \rangle (\rr_T|\rr_S) \left ( {\rm Prob}^{\rightarrow \rightarrow} \langle \psi^{\rightarrow \rightarrow} \rangle + {\rm Prob}^{\rightarrow \leftarrow} \langle \psi^{\rightarrow \leftarrow} \rangle + {\rm Prob}^{\rightarrow \uparrow}  \langle \psi^{\rightarrow \uparrow} \rangle \right )
\end{equation}
The last term can be simplified:
\begin{equation}
\langle \psi \rangle =  {\rm Prob}^{\rightarrow \rightarrow} \langle \psi^{\rightarrow \rightarrow} \rangle + {\rm Prob}^{\rightarrow \leftarrow} \langle \psi^{\rightarrow \leftarrow} \rangle + {\rm Prob}^{\rightarrow \uparrow}  \langle \psi^{\rightarrow \uparrow} \rangle,
\end{equation}
where $\langle \psi \rangle$ is the mean first-passage time for the vacancy, starting from $\rr - \ee_i$ to $\rr$. Using the Kac' formula, we simply obtain $\langle \psi \rangle = 1/P_{\rm stat} = X^d$.

\bigskip

The important point is here that we can only consider the mean first-passage time of the vacancy, without computing the conditional mean first-passage times for each possible jump $\ee_j$. This simplification can be done as soon as the probabilities ${\rm Prob}^{\rightarrow \rightarrow}$ and the conditional mean waiting times $\langle \psi^{\rightarrow \rightarrow} \rangle$ do not depend on the position $\rr$.

\subsubsection{Mean first-passage Time}

We can combine the previous result to obtain the mean first-passage time for a tracer evolving in a crowded periodic euclidian lattice with $n=1$ vacancy:
\begin{equation}
\langle \Tt_t (\rr_S,\rr_T) \rangle  = X^d \langle \Tt(\rr_T|\rr_S) \rangle \label{eq:MFPT1lac}
\end{equation}
The mean first-passage time for a tracer in a crowded environment with $n=1$ vacancy is simply the mean first-passage time of a persistent random walker multiplied by the mean first return time of the vacancy. $\langle \Tt(\rr_T|\rr_S) \rangle$ is here the average first-passage time for a persistent random walk, already obtained previously, that we briefly recall here:
\begin{equation}
\langle \Tt(\rr_T|\rr_S) \rangle = \frac{1}{d}  \sum_{{\bf q}\neq {\bf 0}} \left ( \sum_{\ee_j \in \mathcal{B}} \frac{ 1-(\epsilon + \delta) \cos \left (2 \pi {\bf q}.\ee_i \right )}{1 + \epsilon^2 - 2 \epsilon \cos ( 2 \pi {\bf q}.\ee_i) - \delta^2}  \frac{1 - e^{2 \imath \pi {\bf q}.(\rr_S - \rr_T)}}{\displaystyle 1-2 p_3 \sum_{\ee_j \in \mathcal{B}} \frac{\cos ( 2 \pi {\bf q}.\ee_j )-(\epsilon + \delta)}{1 + \epsilon^2 - 2 \epsilon \cos ( 2 \pi {\bf q}.\ee_j)-\delta^2}} \right ), \label{eq:MFPT}
\end{equation}
where $\epsilon = {\rm Prob}^{\rightarrow \rightarrow} - {\rm Prob}^{\rightarrow \uparrow}$ and $\delta = {\rm Prob}^{\uparrow \rightarrow} - {\rm Prob}^{\rightarrow \leftarrow}$. Those conditional probabilities are obtained using equation (\ref{eq:Probcond}). We will compute the sum $\epsilon + \delta$:
\begin{eqnarray}
\epsilon + \delta & = &  {\rm Prob}^{\rightarrow \rightarrow} - {\rm Prob}^{\rightarrow \uparrow} +  {\rm Prob}^{\uparrow \rightarrow} - {\rm Prob}^{\rightarrow \leftarrow} \nonumber \\
& = &  {\rm Prob}^{\rightarrow \rightarrow} - {\rm Prob}^{\rightarrow \leftarrow} \nonumber \\
& = &  \frac{1}{2d} \Big ( H \left ( {\bf 0} \right ) - H \left ( \ee_1 \right ) +  H \left ( 2 \ee_i \right ) -  H \left ( \ee_1\right ) \Big ) - \frac{1}{2d} \Big ( H \left ( {\bf 0} \right ) - H \left ( \ee_1 \right ) +  H \left ( {\bf 0} \right ) -  H \left ( - \ee_1\right ) \Big ) \nonumber \\
& = &  \frac{1}{2d} \Big ( H \left ( 2 \ee_i \right ) -  H \left ( {\bf 0} \right ) \Big ) \nonumber \\
& = &  \frac{1}{2d \displaystyle \prod_{i=1}^d X_i}  \sum_{{\bf q} \neq {\bf 0}} \frac{\displaystyle e^{4 \imath \pi {\bf q}.\ee_1}- 1}{\displaystyle 1-\frac{1}{d} \sum_{i=1}^d \cos \left ( 2 \pi {\bf q}.\ee_i \right )} = \frac{1}{2d \displaystyle \prod_{i=1}^d X_i}  \sum_{{\bf q} \neq {\bf 0}} \frac{\displaystyle \cos \left ( 4 \pi {\bf q}.\ee_1 \right )- 1}{\displaystyle 1-\frac{1}{d} \sum_{i=1}^d \cos \left ( 2 \pi {\bf q}.\ee_i \right )} \nonumber \\
& = & - \frac{1}{\displaystyle \prod_{i=1}^d X_i}  \sum_{{\bf q} \neq {\bf 0}} \frac{\displaystyle \sin \left ( 2 \pi {\bf q}.\ee_1 \right )^2}{\displaystyle \sum_{i=1}^d \left ( 1 - \cos \left ( 2 \pi {\bf q}.\ee_i \right ) \right )} = - \alpha ( 0 )
\end{eqnarray}
, where $\alpha (0)$ is the notation used by Nakazato-Kitahara \cite{Nakazato:1980}.

\begin{figure}[htb!]
\centering \includegraphics[width = 0.5\linewidth,clip]{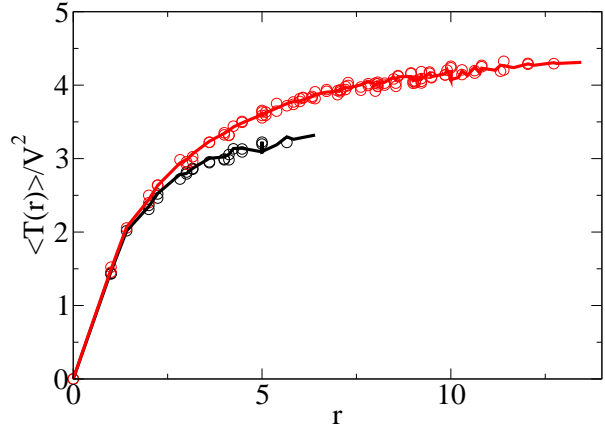} 
\caption{(color online) Mean first-passage time for a tracer evolving on a crowded 2D-periodic Euclidian lattice, with a single vacancy. The black circles stand for the simulation results on a $10\times10$ network, the red circles for a $20 \times 20$ network. The continuous line corresponds to the analytical result of equation (\ref{eq:MFPT1lac}).}
\label{fig:Lacuner1}
\end{figure}

Figure \ref{fig:Lacuner1} shows the mean first-passage time of a tracer diffusing on a crowded 2D periodic Euclidian lattice, as a function of the initial (Euclidian) distance between the tracer and the target ($r = \sqrt{\| \rr_S - \rr_T \|^2}$). The black circles strand for the numerical simulation result on a $10 \times 10$ network, red circles for a $20 \times 20$ network. The continuous line corresponds to the analytical result of equation (\ref{eq:MFPT1lac}). The analytical result fits perfectly the numerical results.

\bigskip

We obtained, for a periodic Euclidian network, an analytical result for the mean first-passage time of a tracer in a crowded environment, with $n=1$ vacancy. Using the result of equation (\ref{eq:MFPT1lac}), we can obtain the GMFPT, with the same formalism as for the persistent random walker.

\break

One can obtain a large volume limit of this mean first-passage time. Equation (\ref{eq:MFPT1lac}) can be simplified using the large volume approximation of \cite{Tejedor:2012}:
\begin{eqnarray}
\langle \Tt_t (\rr_S,\rr_T) \rangle & = & X^d \langle \Tt(\rr_T|\rr_S) \rangle \nonumber \\
 & = & X^d \left ( A \left ( \epsilon , \delta \right ) X^d + \frac{1-\epsilon - \delta}{1+\delta+\epsilon} \left . \langle \Tt(\rr_T|\rr_S) \rangle \right |_0 \right )
\end{eqnarray}
where, for the record:
\begin{equation}
 A \left ( \epsilon , \delta \right ) = \frac{\delta-\epsilon}{1-\epsilon+\delta} + \frac{2 \epsilon}{(1 + \epsilon + \delta) (1 - \epsilon + \delta)} B_d
\end{equation}
and
\begin{equation}
B_d = \underset{X \to \infty}{\lim} \frac{1}{X^d} \sum_{{\bf q} \neq {\bf 0}} \frac{\displaystyle \frac{1}{d} \sum_{{\bf e}_j \in \mathcal{B}} \left ( 1 - \cos \left ( 2 \pi {\bf q}.{\bf e}_j \right ) \right )^2}{\displaystyle \left ( \frac{1}{d} \sum_{{\bf e}_j \in \mathcal{B}} 1 - \cos \left ( 2 \pi {\bf q}.{\bf e}_j \right ) \right )^2}
\end{equation}
Here, we have in 2D the following approximation:
\begin{eqnarray}
\left . \langle \Tt(\rr_T|\rr_S) \rangle \right |_0 \underset{X \to \infty}{\to} X^2 \left ( 1 + \frac{2}{\pi} \ln (r) \right )\\
\epsilon \underset{X \to \infty}{\to}  \frac{1}{4} \frac{1}{(2 \pi)^2} \int_{-\pi}^{\pi} dk_1 \int_{-\pi}^{\pi} dk_2  \frac{\cos \left ( 2 k_1 \right ) - \cos \left ( k_1 + k_2 \right ) }{\displaystyle \sum_{i= 1}^d \left ( 1 - \cos \left ( k_i \right ) \right )} = \frac{3}{\pi} - 1 \\
\delta \underset{X \to \infty}{\to}  \frac{1}{4} \frac{1}{(2 \pi)^2} \int_{-\pi}^{\pi} dk_1 \int_{-\pi}^{\pi} dk_2  \frac{\cos \left ( k_1 + k_2 \right ) - 1 }{\displaystyle \sum_{i= 1}^d \left ( 1 - \cos \left ( k_i \right ) \right )} = -\frac{1}{\pi} \\
\end{eqnarray}

It leads to the following approximation in 2D:
\begin{eqnarray}
\langle \Tt_t \left ( r = \sqrt{ \| \rr_T - \rr_S \|^2} \right ) \rangle  & = & X^4 \left ( \left ( 1 - \frac{\pi}{2}\right ) + \pi \frac{\pi - 3}{2 \pi - 4} B_2 + \frac{2}{\pi-2}\left ( 1 + \frac{2}{\pi} \ln (r) \right )\right ) \label{eq:ApproxT2D}
\end{eqnarray}

\break

\begin{figure}[htb!]
\centering \includegraphics[width = 0.5\linewidth,clip]{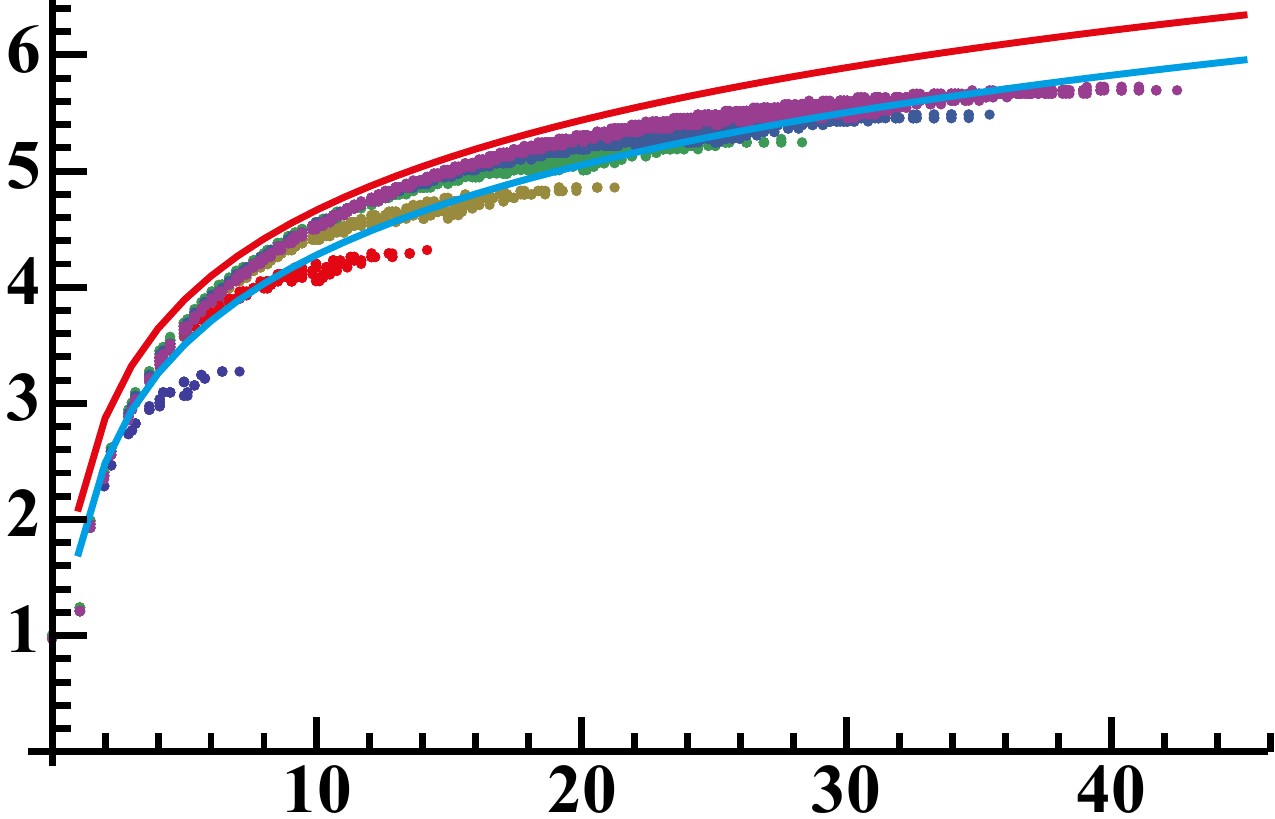} 
\caption{(color online) Mean first passage time divided by $X^{2d}$, $\langle \Tt_t \left ( r \right ) \rangle/X^{2d}$ for a tracer evolving on a 10x10 (blue), 20x20 (red), 30x30 (yellowish), 40x40 (green), 50x50 (blue) or 60x60 (purple) periodic euclidian lattice for an almost compact system (1 vacancy) as a function of the initial distance between tracer and target, compared to Eqn. (\ref{eq:ApproxT2D}) (continuous line).}
\label{fig:approx2D}
\end{figure}

\bigskip

In 3D, we obtain:
\begin{eqnarray}
\left . \langle \Tt(\rr_T|\rr_S) \rangle \right |_0 \underset{X \to \infty}{\to} X^3 \left ( 1 + \frac{3}{2 \pi} \left (1-\frac{1}{r} \right ) \right )\\
\epsilon \underset{X \to \infty}{\to}  \frac{1}{6} \frac{1}{(2 \pi)^3} \int_{-\pi}^{\pi} dk_1 \int_{-\pi}^{\pi} dk_2 \int_{-\pi}^{\pi} dk_3  \frac{\cos \left ( 2 k_1 \right ) - \cos \left ( k_1 + k_2 \right ) }{\displaystyle \sum_{i= 1}^d \left ( 1 - \cos \left ( k_i \right ) \right )} \\
\delta \underset{X \to \infty}{\to}  \frac{1}{6} \frac{1}{(2 \pi)^3} \int_{-\pi}^{\pi} dk_1 \int_{-\pi}^{\pi} dk_2 \int_{-\pi}^{\pi} dk_3  \frac{\cos \left ( k_1 + k_2 \right ) - 1 }{\displaystyle \sum_{i= 1}^d \left ( 1 - \cos \left ( k_i \right ) \right )}
\end{eqnarray}

It leads to the following approximation in 2D:
\begin{equation}
\langle \Tt_t \left ( r = \sqrt{ \| \rr_T - \rr_S \|^2} \right ) \rangle  = X^6 \left ( {\rm Residual} + 1 + \frac{3}{2 \pi} \left ( 1 - \frac{1}{r} \right ) \right ) \label{eq:ApproxT3D}
\end{equation}

\begin{figure}[htb!]
\centering \includegraphics[width = 0.5\linewidth,clip]{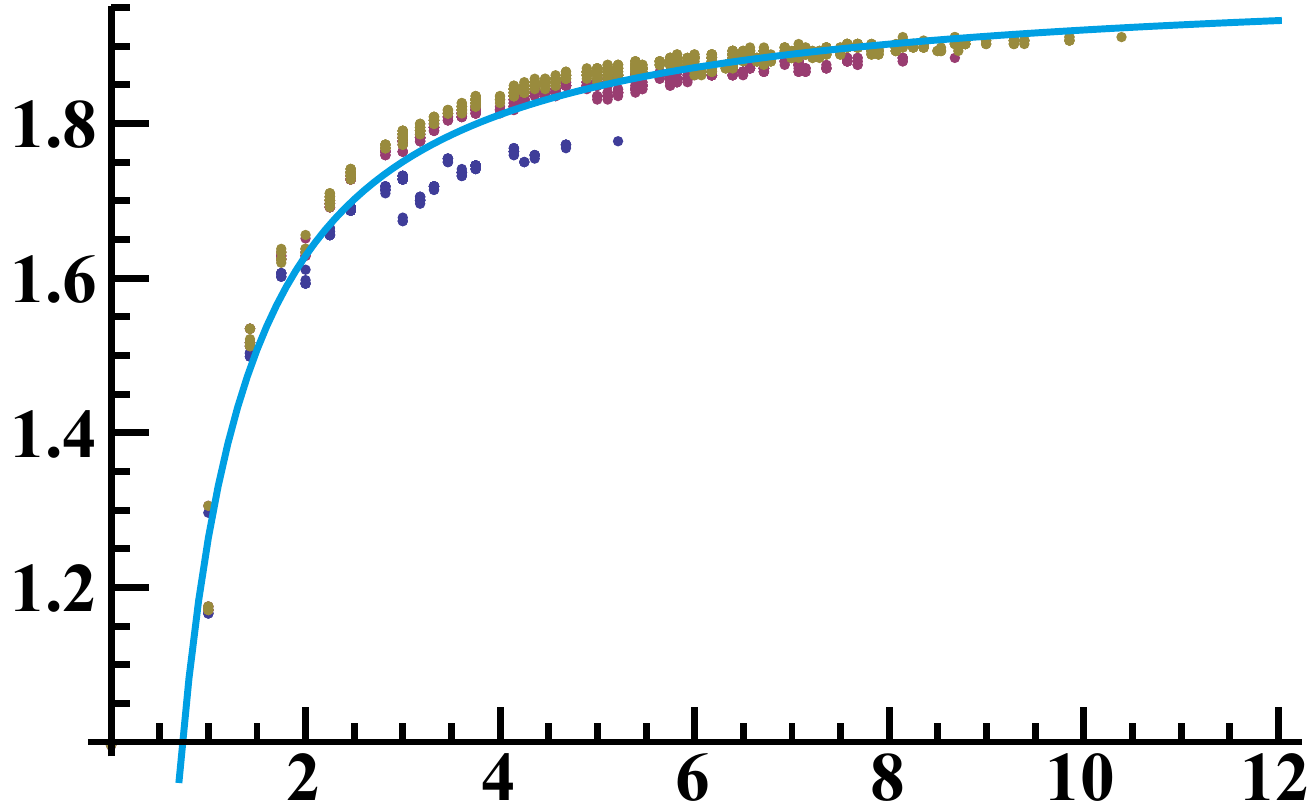} 
\caption{(color online) Mean first passage time divided by $X^{2d}$, $\langle \Tt_t \left ( r \right ) \rangle/X^{2d}$ for a tracer evolving on a 6x6x6 (blue), 10x10x10 (magenta) or 12x12x12 (yellowish) periodic euclidian lattice for an almost compact system (1 vacancy) as a function of the initial distance between tracer and target, compared to Eqn. (\ref{eq:ApproxT3D}) (continuous line).}
\label{fig:approx3D}
\end{figure}

\break

\subsection{Extension for $n \geq 1$}

We will now propose an approximation for the mean first-passage time of a random walker evolving on a crowded periodic euclidean lattice with $n \geq 1$.

The basic idea is to consider that we still have a $n = 1$ vacancy problem, that we can solve analytically, but where the conditional jump probabilities $(\epsilon,\delta)$ depend on $n$. We will apply the result of equation (\ref{eq:MFPT1lac}), with $(\epsilon_n,\delta_n)$, now depending on $n$. The mean waiting time between two tracer steps $\langle \psi \rangle$ will also depends on $n$. Combining those two results, we will be able to propose an approximation of the mean first-passage time of a tracer in a crowded periodic Euclidian lattice for an arbitrary vacancy number $n$.

To compute the influence of $n$, we will rely on the first-passage density, using the approximation of reference \cite{Benichou:2010}, assuming that $d \geq 2$:
\begin{equation}
\begin{array}{l}
\displaystyle \psi_0(t) = \left ( 1 - \frac{X^d}{\rm GMFPT} \right ) \delta \left ( \frac{t}{\rm GMFPT} \right )  + \frac{X^d}{{\rm GMFPT}^2} \exp \left ( -\frac{t}{\rm GMFPT} \right )\\[3mm]
\displaystyle \psi_b(t) = \frac{1}{\rm GMFPT} \exp \left ( -\frac{t}{\rm GMFPT} \right )
\end{array} \label{eq:FPTvac}
\end{equation}
$\delta(x)$ is here a Dirac delta function. The $X^d$ term come from the fact that the first-passage time for the vacancy $0$ is $X^d$, and the GMFPT is the global mean first-passage time to a given point of the network (they are all equal for a periodic network). GMFPT can be expressed as:
\begin{equation}
{\rm GMFPT} = \frac{X^{2 d} }{X^d -1} H({\bf 0})
\end{equation}

\subsubsection{Approximation of $\langle \psi \rangle$}

We now have to obtain $\langle \psi \rangle$, the time between two tracer steps. It does not matter anymore wether the vacancy $0$ or any bulk vacancy hit the tracer first: as shown for the single vacancy case, we just need to have the mean waiting time between two tracer steps. This time can be approximated by the first-passage time of $n$ particles uniformly distributed. We will assume that we have $n$ vacancies having the first-passage density $\psi_b(t)$:
\begin{eqnarray}
\nonumber \langle \psi \rangle & = & \int_0^{\infty} t \left ( \psi_0 (t) \left ( \int_t^{\infty} \psi_b (t') dt' \right )^{n-1} + (n-1) \psi_b (t) \left ( \int_t^{\infty} \psi_b (t') dt' \right )^{n-2} \int_t^{\infty} \psi_0 (t') dt' \right ) dt\\
\nonumber & = &  \int_0^{\infty} \frac{n X^d}{\rm GMFPT^2} \exp \left ( - \frac{n t}{\rm GMFPT} \right ) t dt \\
 \langle \psi \rangle & = & \frac{\rm X^d}{n}
\end{eqnarray}

This result is coherent with the two limit regimes: when $n \to X^d$, $\langle \psi \rangle \to 1$, this time being the one between two step for a random walker without any obstacle. When $n = 1$, $\langle \psi \rangle = X^d$, this time being the mean first return time that can be retreived by the Kac' formula.

\begin{figure}[htb!]
\centering \includegraphics[width = 0.5\linewidth,clip]{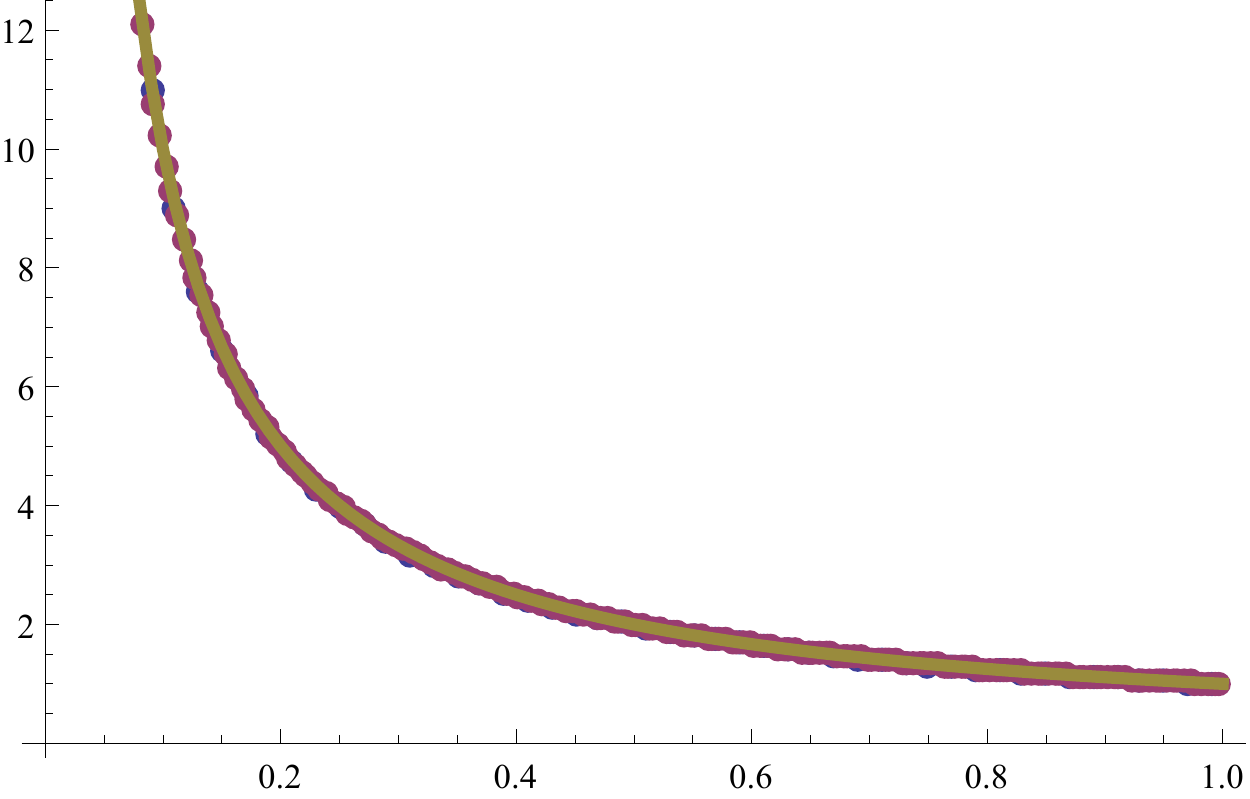} 
\caption{(color online) Mean first waiting time, $\langle \psi \rangle / X^d$ for a tracer evolving on a 10x10 (blue) or 20x20 (red) periodic euclidian lattice as a function of the vacancy density ($n/X^d$), compare to $f(x) = 1/x$ (yellowish line).}
\label{fig:FRT}
\end{figure}

\begin{figure}[htb!]
\centering \includegraphics[width = 0.5\linewidth,clip]{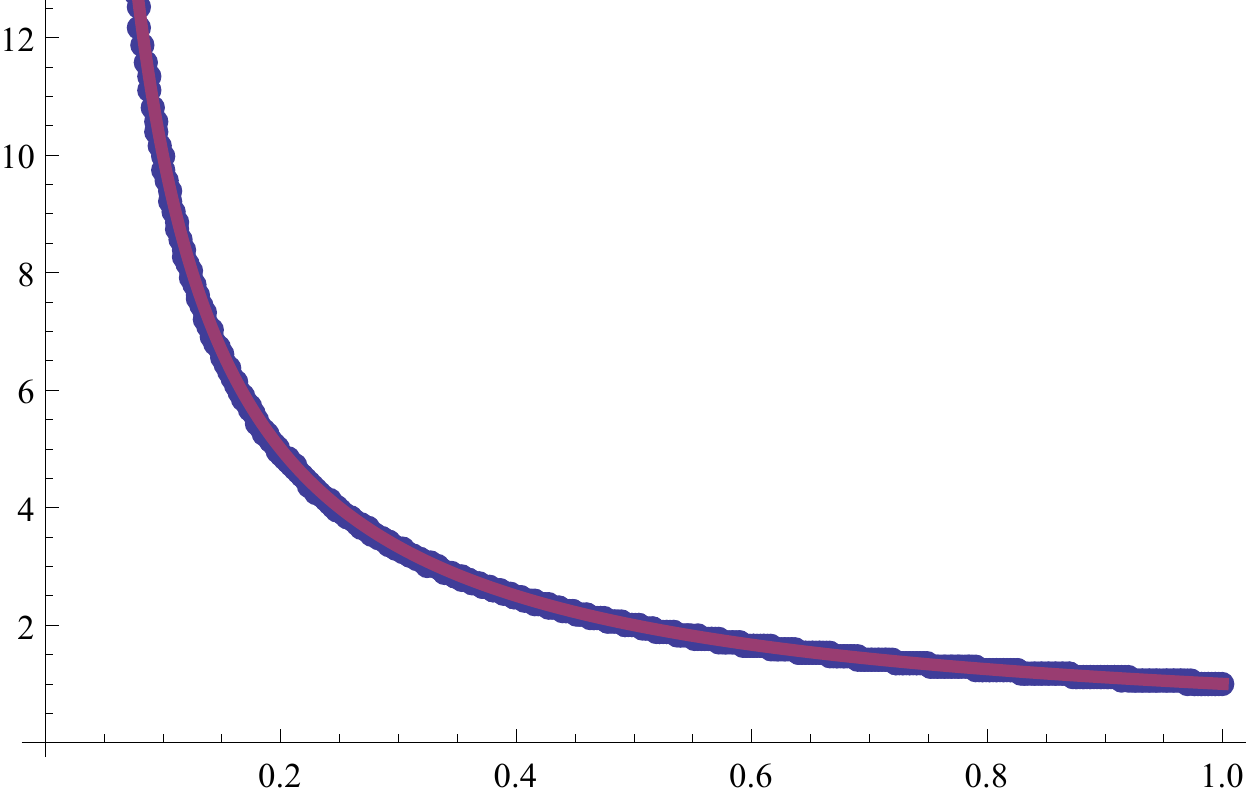} 
\caption{(color online) Mean first waiting time, $\langle \psi \rangle / X^d$ for a tracer evolving on a 10x10x10 periodic euclidian lattice (blue) as a function of the vacancy density ($n/X^d$), compare to $f(x) = 1/x$ (red line).}
\label{fig:FRT3D}
\end{figure}

We can check this result through numerical simulation, as shown in Figure \ref{fig:FRT} for 2D lattices, and in Figure \ref{fig:FRT3D} for 3D lattices.

\break

This result can be retrieved by a mean field approximation. At each step, the tracer can move if the targeted site is free, namely occupied by a vacancy. If there is $n$ vacancies in a networks of size $X^d$, the probability to make a step at time $t$ is:
\begin{equation}
{\rm Prob}(\psi = t) = \frac{n}{X^d} \left ( 1 - \frac{n}{X^d} \right )^{t-1}
\end{equation}
One can quickly obtain the average waiting time:
\begin{equation}
\langle \psi \rangle = \sum_{t=0}^{\infty} t {\rm Prob}(\psi = t) =  \sum_{t=0}^{\infty} t \frac{n}{X^d} \left ( 1 - \frac{n}{X^d} \right )^{t-1} = \frac{X^d}{n}
\end{equation}

\subsubsection{Estimation of $\epsilon_n$ and $\delta_n$}

When the tracer evolves with $n$ vacancies, we will consider that $n-1$ ``bulk'' vacancies are uniformly distributed in all space, while the $0$ vacancy is right next to the tracer. The persistance will be approximated as follow: (i) there is a persistance if, and only if, the $0$ vacancy hit the tracer first, (ii) short term effects are predominant for persistance, namely there is a persistance only if the $0$ vacancy hit the tracer after few steps. These two effects are approximated with the following expression:
\begin{equation}
\epsilon_n + \delta_n = \left ( 1 - \frac{n}{X^d} \right ) \left ( \epsilon + \delta \right )
\label{Persistance}
\end{equation}

\break

The first parenthesis in the right hand side is the probability that no ``bulk'' vacancy hit the tracer during the first steps (in fact during the first step, with if needed a re-scaled time unit), and the second parenthesis the persistance related to the vacancy $0$ if left alone.

\begin{figure}[htb!]
\centering \includegraphics[width = 0.5\linewidth,clip]{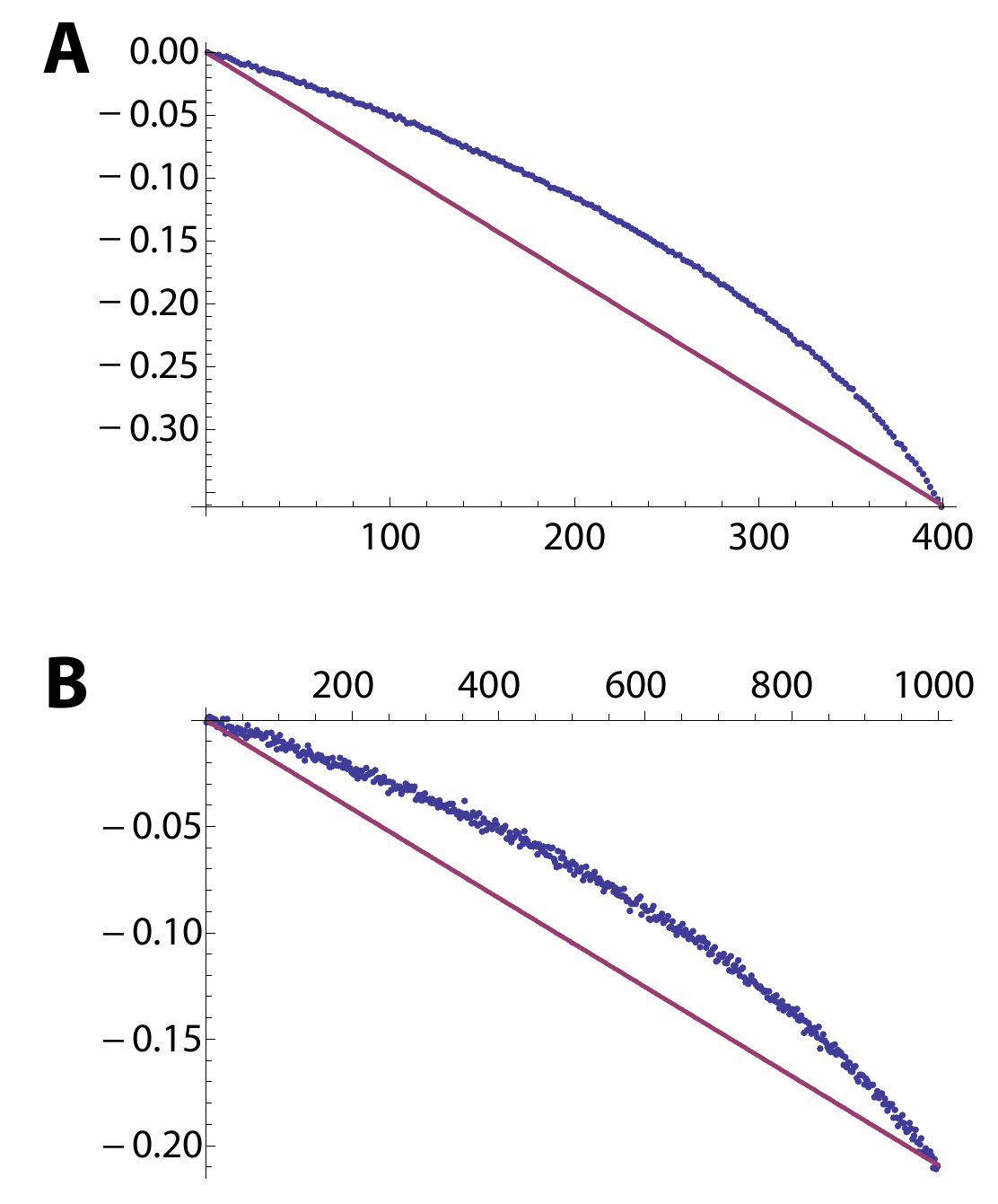} 
\caption{(color online) Plot of the simulated persistance $\epsilon_n+\delta_n$ for a tracer evolving on a periodic euclidian lattice as a function of the obstacle number $X^d-n$ (blue dots) compared to Eq. (\ref{Persistance}) (purple line). {\bf A} is for a 20x20 lattice, and {\bf B} for a 10x10x10 lattice.}
\label{fig:Persistance}
\end{figure}

This assumption can be tested against numerical simulation, and appears to be rough, but give the correct order of magnitude, as shown in Figure \ref{fig:Persistance}.

One can note that this expression implicitely takes into account the hard sphere effect: a simple computation with $n$ freely non interacting vacancy should give a factor $1/n$ instead of $n/X^d$.

\subsubsection{First-passage time with $n$ vacancies}

At last, the first-passage time with $n$ vacancies is the one of a persistent random walk, with conditional probabilities $(\epsilon_n,\delta_n)$ and a mean waiting time $\langle \psi \rangle = X^d/n$:
\begin{eqnarray}
\nonumber \langle \Tt_t (\rr_S,\rr_T,n) \rangle & = & \frac{X^d}{n} \left ( \frac{X^d(\delta_n-\epsilon_n)}{1-\epsilon_n + \delta_n} + \frac{1+\epsilon_n^2-\delta_n^2}{(1+\epsilon_n+\delta_n)(1-\epsilon_n+\delta_n)} \right .\\
& & \hspace{0cm} \times \left .  \sum_{{\bf q}\neq {\bf 0}}\frac{1 - e^{2 \imath \pi {\bf q}.(\rr_S - \rr_T)}}{\displaystyle 1-\frac{(\epsilon_n-1)^2-\delta_n^2}{d} \sum_{\ee_j \in \mathcal{B}} \frac{\cos ( 2 \pi {\bf q}.\ee_j )}{1 + \epsilon_n^2 - 2 \epsilon_n \cos ( 2 \pi {\bf q}.\ee_j)-\delta_n^2}} \right ) \label{eq:nlac}
\end{eqnarray}
with
\begin{equation}
\begin{array}{l}
\displaystyle \epsilon_n \simeq  \left ( 1 - \frac{n}{X^d} \right ) \epsilon\\[3mm]
\displaystyle \delta_n \simeq \left ( 1 - \frac{n}{X^d} \right ) \delta
\end{array}
\end{equation}

\break

One can compare this result with the Nakazato-Kitahara approximation:
\begin{equation}
\langle \Tt_t (\rr_S,\rr_T,n) \rangle  =  \frac{1}{D_{NK}(n)} \langle \Tt_t (\rr_S,\rr_T,n=X^d-1) \rangle) \label{eq:NK}
\end{equation}
with
\begin{equation}
D_{NK}(n) = (1-c) \frac{1-\alpha (0)}{\displaystyle 1 - \alpha (0) \frac{2-3 c}{2 - c}} = \frac{n}{X^d} \frac{1+\epsilon+\delta}{\displaystyle 1+\left ( \epsilon+\delta \right ) \frac{3 \frac{n}{X^d} - 1}{1+\frac{n}{X^d}}}
\end{equation}

One can transform both our approximation and Nakazato-Kitahara' using the large volume limit, as already explained for the case with 1 vacancy:
\begin{eqnarray}
\langle \Tt_t (r,n) \rangle & = & \frac{X^{2 d}}{n} \left ( \frac{(\delta_n-\epsilon_n)}{1-\epsilon_n + \delta_n} + \frac{(2 \epsilon_n)}{(1+\epsilon_n + \delta_n) (1-\epsilon_n + \delta_n)} B_d +  \frac{1-\epsilon_n-\delta_n}{1+\epsilon_n+\delta_n} \Big ( G(0) - G(r) \Big ) \right ) \label{eq:nlacapprox}\\
\langle \Tt_t (r,n) \rangle_{NK} & = & \frac{X^{2 d}}{n} \frac{1 + \left ( \epsilon + \delta \right ) \frac{\displaystyle 3 n - X^d}{\displaystyle n + X^d}}{1+\epsilon + \delta} \Big ( G(0) - G(r) \Big )  \label{eq:nlacapproxNK}
\end{eqnarray}

\begin{figure}[htb!]
\centering \includegraphics[width = 0.5\linewidth,clip]{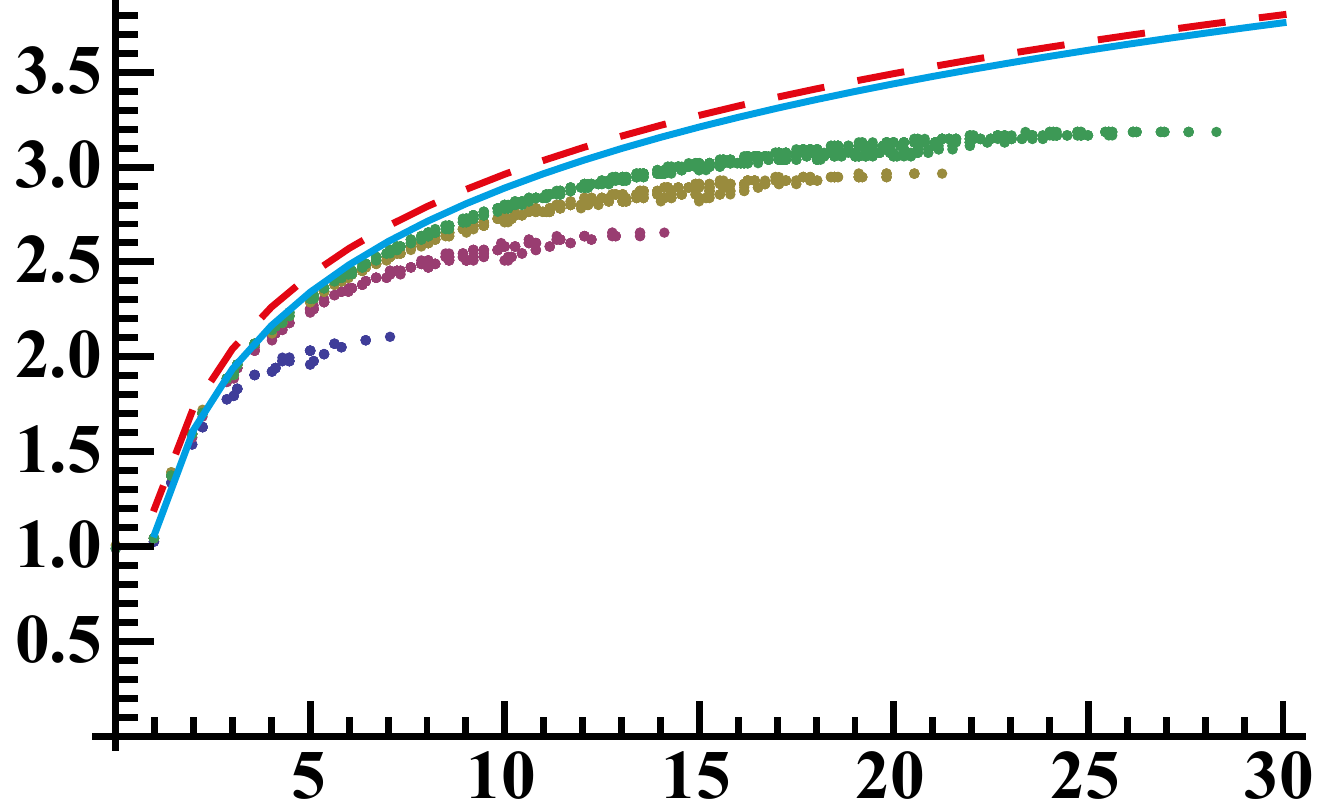} 
\caption{(color online) Mean first-passage time for a tracer evolving on a 10x10 (blue dots), 20x20 (magenta dots), 30x30 (yellowish dots) and 40x40 (green dots) periodic euclidian lattice, for an obstacle concentration of 30 \%, rescaled by $X^{2d}/n$, as a function of the distance $r$ between the source and the target. The continuous blue lines stand for the approximation of equation (\ref{eq:nlacapprox}), and the dashed red line for the classical approximation of Nakazato-Kitahara of equation (\ref{eq:nlacapproxNK}).}
\label{fig:FPT2D-30p}
\end{figure}

\begin{figure}[htb!]
\centering \includegraphics[width = 0.5\linewidth,clip]{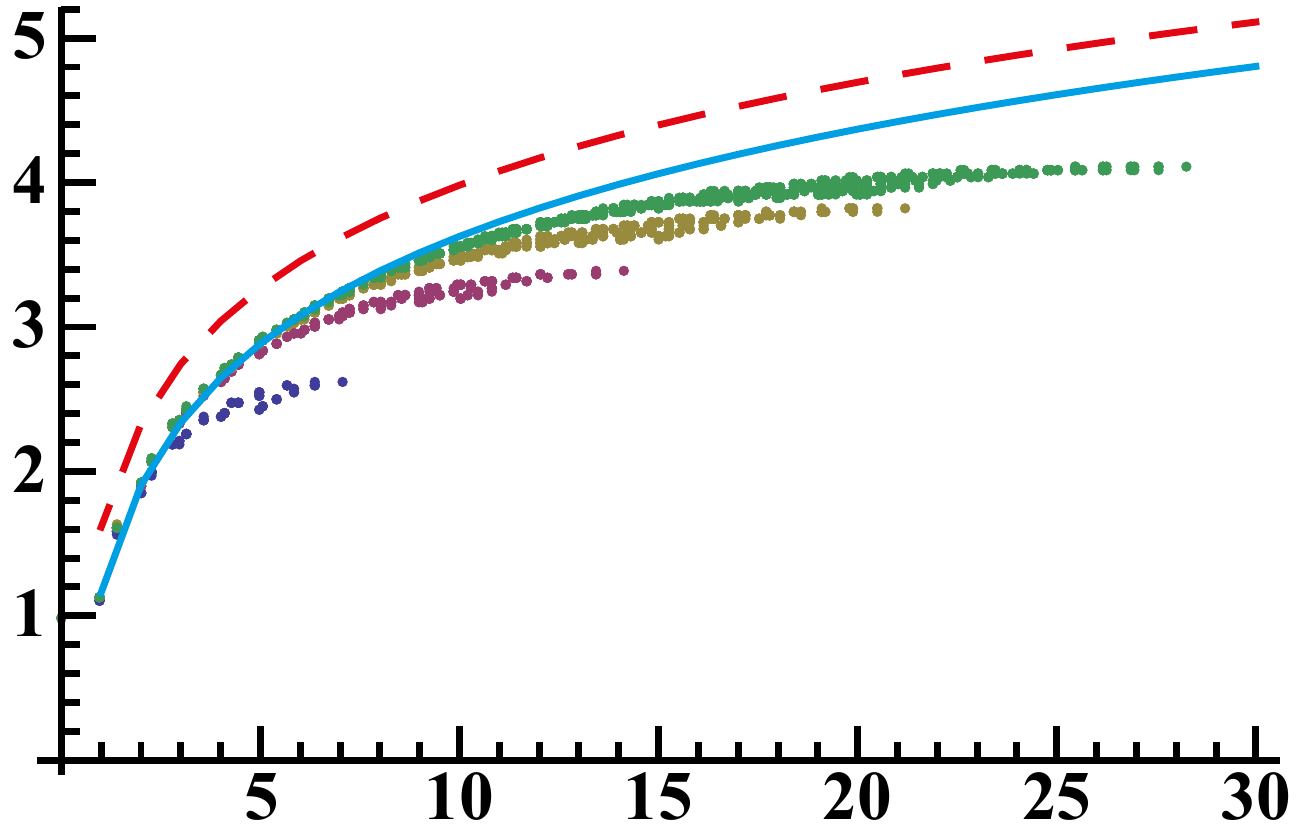} 
\caption{(color online) Mean first-passage time for a tracer evolving on a 10x10 (blue dots), 20x20 (magenta dots), 30x30 (yellowish dots) and 40x40 (green dots) periodic euclidian lattice, for an obstacle concentration of 70 \%, rescaled by $X^{2d}/n$, as a function of the distance $r$ between the source and the target. The continuous blue lines stand for the approximation of equation (\ref{eq:nlacapprox}), and the dashed red line for the classical approximation of Nakazato-Kitahara of equation (\ref{eq:nlacapproxNK}).}
\label{fig:FPT2D-70p}
\end{figure}

The prefactor is slightly different between the two approximations, but the main difference is that the Nakazato-Kitahara approximation does not contain any constant value before the $G(0)-G(r)$ term, the so-called ``residual'' mean first-passage time. As shown for an obstacle concentration of 30 \% (Fig. \ref{fig:FPT2D-30p}) or 70 \% (Fig. \ref{fig:FPT2D-70p}), when the crowding increases, the residual MFPT is clearer, and the Nakazato-Kitahara approximation becomes less efficient. When the obstacle number goes to $X^d$, one find back the $n=1$ limit when our approximation is exact.

\begin{figure}[htb!]
\centering \includegraphics[width = 0.95\linewidth,clip]{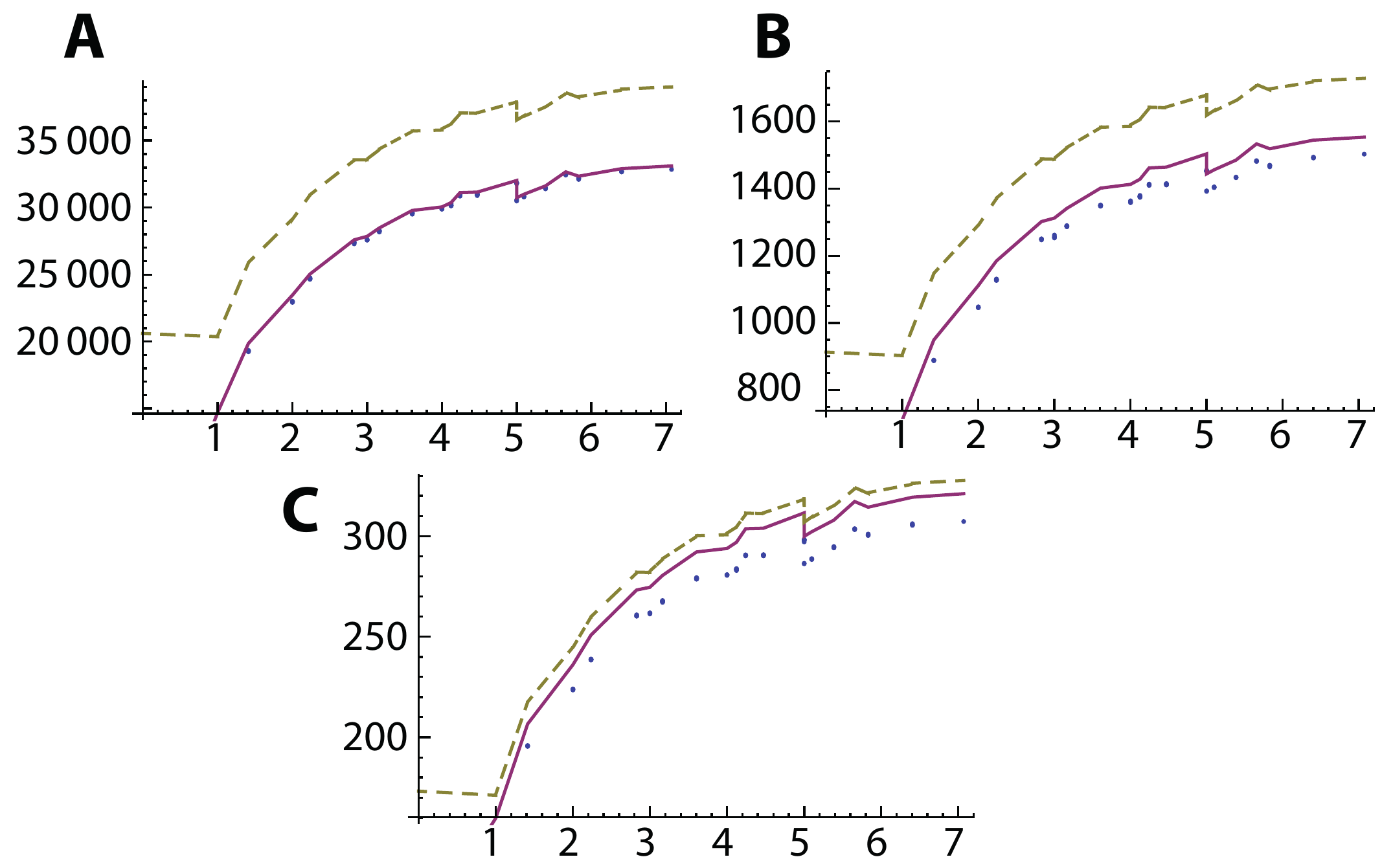} 
\caption{(color online) Mean first-passage time for a tracer evolving on a 10x10 periodic euclidian lattice. The blue dots stand for numerical simulations, the continuous purple lines stand for the approximation of equation (\ref{eq:nlac}), and the dashed yellowish line for the classical approximation of Nakazato-Kitahara of equation (\ref{eq:NK}). A is 99 obstacles (or 1 vacancy), B for 81 obtacles (or 19 vacancies), and C for 31 obstacles (69 vacancies).}
\label{fig:FPT2D}
\end{figure}

Figure \ref{fig:FPT2D} shows the mean first-passage time for a tracer evolving on a $10 \times 10$ periodic Euclidian lattice, as a function of the initial distance between the tracer and the target. The circles stand for numerical simulations, for three crowding conditions. The continuous purple lines stand for the approximation of equation (\ref{eq:nlac}), and the dashed yellowish line for the classical approximation fo equation (\ref{eq:NK}): one can clearly see that our approximation fits very well the simulation results, and performs better that the classical approximation.

\begin{figure}[htb!]
\centering \includegraphics[width = 0.5\linewidth,clip]{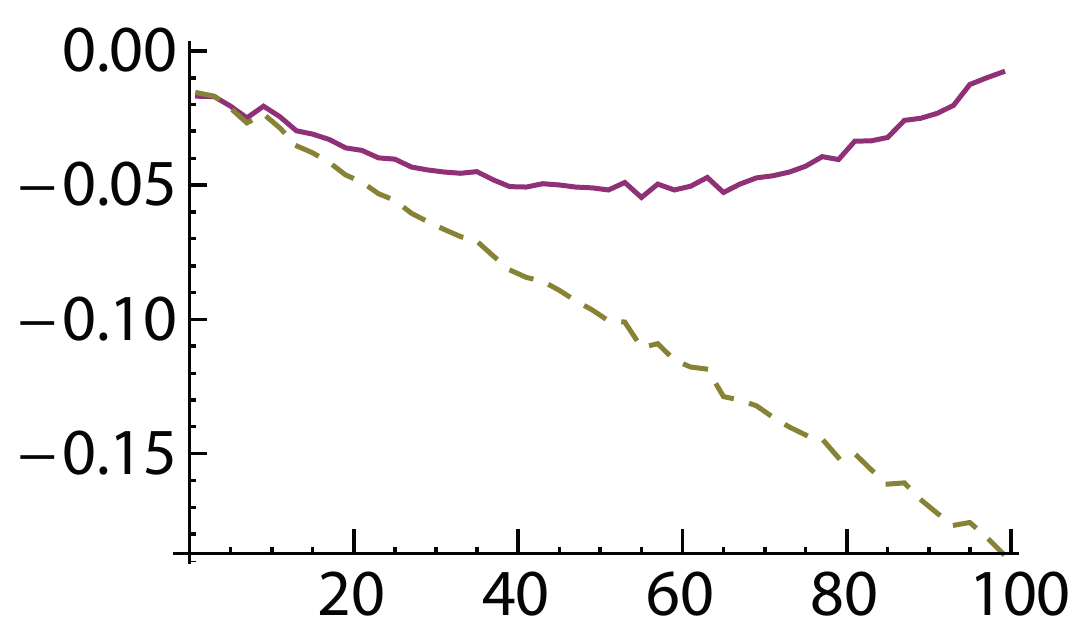} 
\caption{(color online) Relative error on the mean first-passage time for a tracer evolving on a 10x10 periodic euclidian lattice (difference between simulation and prediction, divided by the simulation value) for a target located in $(X/2,X/2)$ (worst case), as a function of the obstacle number. The continuous purple lines stand for the approximation of equation (\ref{eq:nlac}), and the dashed yellowish line for the classical approximation of Nakazato-Kitahara of equation (\ref{eq:NK}).}
\label{fig:FPT2Derror}
\end{figure}

Our approximation is exact for a very crowded system (1 vacancy), and takes somehow into account the hard sphere effect in the $\epsilon_n$ and $\delta_n$ approximation. As shown in Figure \ref{fig:FPT2Derror}, the relative error is always below 5 \% for our approximation, while the classical approximation exceed 15 \% for very crowded systems.

\break

\section*{Conclusion}

We have developed here a quite simple approximation of hard-core crowding on discrete network. This model was developed to see what would be the effect of crowding on first-passage processes: we surprisingly retrieve a ``simple'' persistent random walk with a waiting time at each step. This process shares some features with CTRW at short time, in particular a scattering of the diffusion coefficient due to the variability of the vacancy first return time, and become a persistent random walk with an effective diffusion coefficient at longer time.

\break

The final approximation is astonishingly good: we consider a tracer diffusing on a network with many hard-core obstacles interacting, and we see that the mean first-passage time can be approximated very decently through a rather simple expression. The absence of correlation between conditional probabilities and conditional mean first-passage time is surprising, and simplifies greatly the formalism.

\bibliographystyle{plain}

\bibliography{../../biblio}
\end{document}